\documentclass{emulateapj}
\usepackage{apjfonts}

\input{iondefs.sty}

\slugcomment{{\it Accepted}: Astrophysical Journal (12/06)}

\shorttitle{\sc Heterogeneity of Galaxy Halos}
\shortauthors{\sc Churchill {\etal}}

\begin{document}


\title{On the Heterogeneity of Metal--Line and {\Lya} Absorption in
Galaxy ``Halos'' at $\lowercase{z}\sim 0.7$}


\author{\sc
Christopher W. Churchill\altaffilmark{1},  
Glenn G. Kacprzak\altaffilmark{1}, \\ 
Charles C. Steidel\altaffilmark{2},
and
Jessica L. Evans\altaffilmark{1}
}

\altaffiltext{1}{New Mexico State University, Las Cruces, NM 88003
{\tt cwc, glennk, jlevans@nmsu.edu}}

\altaffiltext{2}{Caltech, Pasadena, CA 91125 
{\tt ccs@astro.caltech.edu}}

\begin{abstract}

We examine the properties of two galaxy ``halos'' at $z \sim 0.7$ in
the TON 153 ($z_{\rm em} = 1.01$) quasar field.  The first
absorber--galaxy pair (G1) is a $z=0.672$, $L_{B} = 4.3~L^{\ast}_{B}$,
E/S0 galaxy probed at $D=58$ kpc.  G1 is associated with a remarkable
five--component {\Lya} complex having $\tau _{_{\rm LL}} \leq 0.4 $,
$W_r({\Lya}) = 2.8$~{\AA}, and a velocity spread of $\Delta v =
1420$~{\kms}.  We find no {\MgII}, {\CIV}, {\NV}, nor {\OVI}
absorption in these clouds and infer metallicity upper limits of $-3
\leq \log Z/Z_{\odot} \leq -1$, depending upon assumptions of
photoionized or collisionally ionized gas.  The second
absorber--galaxy pair (G2) is a $z=0.661$, $L_{B} = 1.8~L^{\ast}_{B}$,
Sab galaxy probed at $D=103$ kpc.  G2 is associated with
metal--enriched ($\log Z/Z_{\odot} \simeq -0.4$) photoionized gas having
$N({\HI}) \simeq 18.3$ [{\cmsq}] and a velocity spread of $\Delta v =
200$~{\kms}.  The very different G1 and G2 systems both have
gas--galaxy properties inconsistent with the standard luminosity
dependent galaxy ``halo'' model commonly invoked for quasar absorption
line surveys.  We emphasize that mounting evidence is revealing that
extended galactic gaseous envelopes in the regime of $D \leq
100$~kpc do not exhibit a level of homogeneity supporting a
standardized halo model.  Selection effects may have played a central
role in the development of a simple model.  We discuss the G1 and G2
systems in the context of $\Lambda$CDM models of galaxy formation and
suggest that the heterogeneous properties of absorber--galaxy pairs is
likely related to the range of overdensities from which galaxies and
gas structures arise.

\end{abstract}



\keywords{galaxies: halos, evolution --- quasars: individual (TON 153)}

\section{Introduction}
\label{sec:intro}

Observations of baryonic gas in galaxy halos constrain galaxy
evolution models, in particular the successful $\Lambda$CDM model of
structure formation.  However, important quandaries persist, including
the overcooling \citep[e.g.,][]{wr78} and baryon fraction problems
\citep[e.g.,][]{mo02}, in which the cooling of baryonic gas is too
efficient, resulting in compact halos and highly peaked rotation
curves inconsistent with the Tully--Fisher relationship \citep{mmw98}.
Samples of galaxy--absorber pairs in quasar absorption line surveys of
{\Lya} $\lambda 1215$ \citep{lanzetta95,chen98,chen01b}, {\MgIIdblt}
\citep{bb91,lebrun93,sdp94,steidel97,cwc-china}, {\CIVdblt}
\citep{chen01a}, and {\OVIdblt}
\citep{sembach04,tripp-china,tumlinson05,tripp06} absorption provide
key data for addressing the overarching question-- what processes
regulate and sustain the baryon fraction and kinematic, chemical and
ionization conditions of galaxy halos in the context of $\Lambda$CDM
models?

The gas boundaries of ``absorption selected'' galaxies are
characterized using the Holmberg--like \citep{holmberg75}
relationship,
\begin{equation}
R(L) = R^{\ast} \left( \frac{L_B}{L_{B}^{\ast}} \right) ^{\beta} ,
\label{eq:rlb}
\end{equation}
where $R^{\ast}$ is the halo size for an $L^{\ast}$ galaxy, which
scales as $R^{\ast} \propto [(dN/dz)/(C_f f_{gal} {\cal K} )]^{1/2}$,
where $dN/dz$ is the absorption redshift path density, $C_f$ is the
covering factor, $f_{gal}$ is the fraction of galaxies with absorbing
halos, and ${\cal K}$ is the geometry factor (${\cal K} = 0.5$ for
thin disks; ${\cal K} = 1$ for spherical).  Estimates of $R^{\ast}$
and $\beta$ apply to surveys conducted to a fixed absorption
sensitivity for a given ion/transition and employ the galaxy
luminosity function \citep[see][]{lanzetta95}.

For {\MgII} absorbers with $W_{r}(2796) \geq 0.3$~{\AA},
\citet{steidel95} reported $R^{\ast} = 55$~kpc\footnote{Throughout
this paper, we adopt a $H_0=70$~{\kms}~Mpc$^{-1}$, $\Omega_m=0.3$, and
$\Omega _{\Lambda}=0.7$ cosmology.  We have converted all cosmology
dependent physical quantities taken from the literature to this
cosmology, unless explicitly stated.}, $\beta = 0.2$, and $C_f f_{gal}
{\cal K} = 1$.  Lyman limit systems (LLS, $\tau_{_{\rm LL}} \geq 1$)
show one--to--one correspondence with $W_{r}(2796) \geq 0.3$~{\AA}
{\MgII} absorption \citep{archiveI} and have virtually identical
$dN/dz$ \citep{eric-lls95}.  Statistically, LLS have $R^{\ast} \simeq
55$~kpc for an {\it assumed\/} $C_f f_{gal} {\cal K} = 1$.  For {\Lya}
absorption with $W_r({\Lya}) \geq 0.30$~{\AA}, \citet{chen98,chen01b}
reported $C_f f_{gal} {\cal K} = 1$ within $R^{\ast}= 310$~kpc with
$\beta = 0.39$.  For {\CIV} systems with $W_{r}(1548) \geq
0.17$~{\AA}, \citet{chen01a} found $R^{\ast} = 170$~kpc and $\beta =
0.5$, and deduce $C_f f_{gal} {\cal K} = 1$.  There are few
constraints for {\OVI} absorption, which may commonly arise in the
warm--hot intergalactic medium \citep[][however, see Tumlinson {\etal}
2005; Tripp {\etal} 2006]{dave98,dave01}.

\begin{figure*}[bth]
\epsscale{1.0}
\plotone{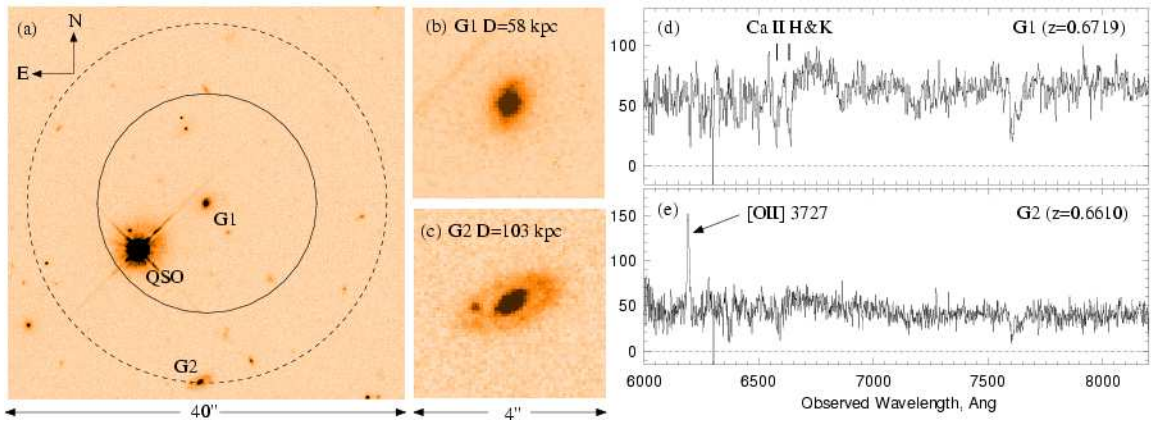}
\caption{(a) A $40{\arcsec} \times 40{\arcsec}$ section of the F702W
WFPC--2 image of the TON 153 field centered on galaxy G1.  The solid
circle represents a $R(L) = 73$ kpc halo for $W_{r}(2796) \geq
0.3$~{\AA} and for LLS absorption. The dashed circle represents a
$R(L) = 123$ kpc halo for $W_{r}(2796) \geq 0.02$~{\AA}. --- (b,c)
$4{\arcsec} \times 4{\arcsec}$ images of G1 and G2.  --- (d,e) The
Keck/LRIS spectra of G1 and G2.}
\label{fig:image}
\end{figure*}

A possible source of bias toward deducing $C_f f_{gal} {\cal K} = 1$,
is that the presence of a gas ``halo'' is requisite for the inclusion
of an absorber--galaxy pair in an absorption selected sample.
Establishing the absorption {\it ex post facto\/} galaxy
identifications has yielded $C_f f_{gal} {\cal K} < 1$;
\citet{bechtold92} reported $C_f f_{gal} {\cal K} \simeq 0.25$ for
$W_{r}(2796) \geq 0.26$~{\AA} ($3~\sigma$) with $D \leq 85$ kpc and
\citet{tripp-china} reported $C_f f_{gal} {\cal K} \sim 0.5$ for
$W_{r}(2796) \geq 0.15$~{\AA} ($3~\sigma$) with $D \leq 50$ kpc.  For
$W_{r}(2796) \geq 0.02$~{\AA} ($5~\sigma$), \citet{cwc-china} reported
that very weak {\MgII} absorption arises both well {\it inside and
outside\/} the $R(L)$ boundary of bright galaxies.  Churchill {\etal}
also find $W_r(2796) > 1$~{\AA} absorption out to $\simeq 2 R(L)$ in
several cases.  Since $\tau _{_{\rm LL}} < 1$ {\HI} traces weak
{\MgII} and $\tau _{_{\rm LL}} > 1$ {\HI} traces strong {\MgII}
\citep{archiveI}, it is clear that $N({\HI})$ is patchy over 4--5
decades for the full range of impact parameters.  These results
clearly challenge the idea of a ``conventional'' galaxy halo
encompassed by Equation~\ref{eq:rlb}.

In this {\it Letter}, we examine two very different absorber--galaxy
pairs at $z\sim 0.7$ in the field of \object{TON 153} (Q$1317+274$)
that both clearly depart from the ``conventional halo'' model deduced
from absorption line surveys.  In \S~\ref{sec:data}, we present the
data.  We examine the galaxy and absorption properties in
\S~\ref{sec:results}.  In \ref{sec:discuss}, we discuss these
contrasting properties in the context of $\Lambda$CDM formation.
Concluding remarks are given in \S~\ref{sec:conclude}.

\section{Data and Analysis}
\label{sec:data}

An F702W image (PID 5984; PI Steidel) of the TON 153 field was
obtained with WFPC--2 facility on board {\it Hubble Space Telescope\/}
({\it HST\/}).  The image was reduced and calibrated using the WFPC--2
Associations Science Products Pipeline (WASPP\footnote{Developed by
the Canadian Astronomy Data Centre (CADC) and the Space
Telescope--European Coordinating Facility (ST--ECF): {\it
http://archive.stsci.edu/hst/wfpc2/pipeline.html}}).  The galaxy
photometry was performed using the Source Extractor (Sextractor)
package \citep{bertin96}, and the $K$--correction was computed using
the formalism of \citet{kim96}.  We used the GIM2D software
\citep{simard02} to obtain quantified morphological parameters of the
galaxies.

The galaxy spectra were obtained with LRIS \citep{oke95} on Keck I in
1999--March.  A $1{\arcsec}$ wide, long slit was used with a 600 lines
mm$^{-1}$ grating blazed at 7500~{\AA}, yielding spectral coverage
from 5770--8340~{\AA} and a resolution of $\hbox{\sc fwhm} \simeq
4.5$~{\AA}.  The LRIS spectra were reduced with IRAF\footnote{IRAF is
written and supported by the IRAF programming group at the National
Optical Astronomy Observatories (NOAO) in Tucson, Arizona. NOAO is
operated by the Association of Universities for Research in Astronomy
(AURA), Inc.\ under cooperative agreement with the National Science
Foundation.}  using the standard NOAO {\it onedspec\/} tasks.  The
optical quasar spectrum was obtained with HIRES \citep{vogt94} on the
Keck I telescope in 1995--January.  Reduction and calibration details
can be found in \citet{cv01}, and \citet{cvc03}.  The UV quasar
spectra were obtained with both G160L/G190H/FOS (PID 2424; PI Bahcall)
and E230M/STIS (PID 8672; PI Churchill) on board {\it HST}. ~The
resolving powers are $R = 170$, $1300$, and $30,000$, respectively.
Reduction and calibration details of the FOS spectra can be found in
\citet{archiveI} and of the STIS spectrum in \citet{ding05}.  All
subsequent analyses of the spectra were then performed using our own
suite of graphical interactive software \citep[following the work
of][]{weakI,archiveI,cv01} for continuum fitting, identifying
features, and measuring absorption feature properties.

\section{Results}
\label{sec:results}

In Figure~\ref{fig:image}$a$, we present a $40{\arcsec} \times
40{\arcsec}$ subsection of the WFPC--2/F702W image of the TON 153
field.  The galaxy G1 is centered in the image and lies at an impact
parameter of $D=58.4$~kpc from the quasar.  Galaxy G2 lies at $D=103$
kpc.  In Figures~\ref{fig:image}$b$ and $c$, we present $4{\arcsec}
\times 4{\arcsec}$ images of the galaxies.  The LRIS spectra of G1 and
G2 are presented in Figures~\ref{fig:image}$d$ and $e$.  A redshift of
$z_{_{\rm G1}}=0.6719$ was determined by Gaussian fitting to the
{\CaII} features and $z_{_{\rm G2}}=0.6610$ was determined by Gaussian
fitting to the strong [{\OII}] $\lambda 3727$ emission.  G1 has a
luminosity of $L_B = 4.3 L_B^{\ast}$. The bulge--to--total ratio is
$B/T = 0.41$.  From the GIM2D modeling, the $C$--$A$ morphological
classification is E/S0 \citep[see][]{abraham96}.  G2 has $L_B = 1.8
L_B^{\ast}$, $B/T=0.81$, and a $C$--$A$ morphological classification
of Sab.

\begin{figure*}[tbh]
\epsscale{1.0}
\plotone{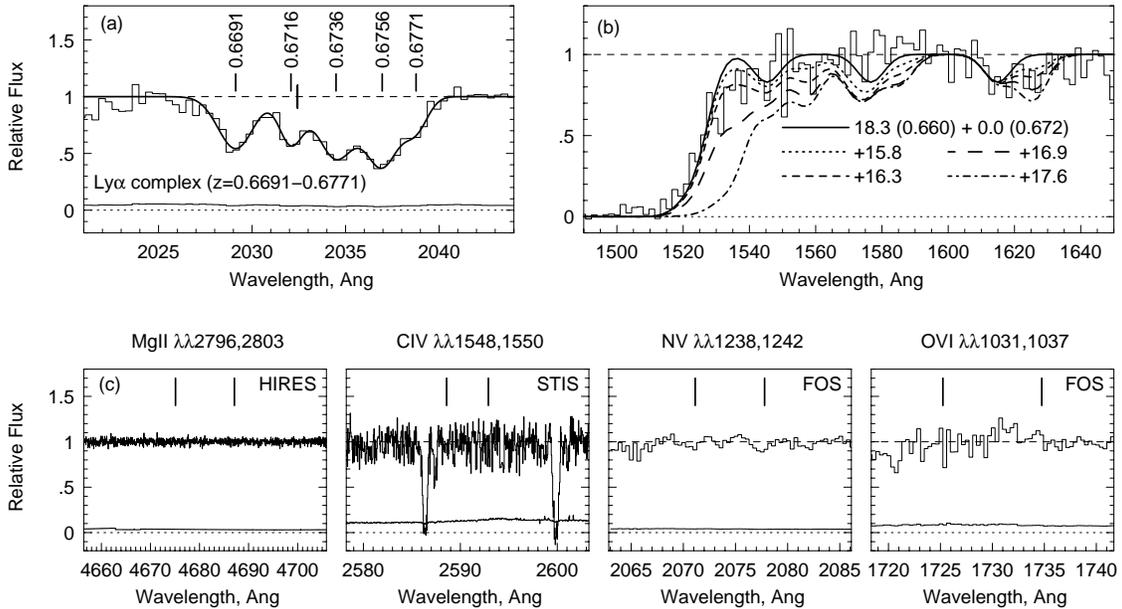}
\caption{(a) The {\Lya} complex (FOS/G190H) spanning $0.6691 \leq z
\leq 0.6771$.  The lowered tick corresponds to $z_{_{\rm
G1}}=0.6719$. --- (b) The Lyman limit region (FOS/G160L), illustrating
our modeling (see text) of the break for the {\Lya} complex in panel
(c). --- (c) Regions of anticipated {\MgII} (HIRES), {\CIV}
(STIS/E230M), and {\NV} and {\OVI} (FOS/G190H) doublet absorption at
the redshift of G1, covering $\pm 1200$~{\kms} (rest--frame) with
respect to $z=0.6719$, or $0.6652 \leq z \leq 0.6785$. The ticks
provide the expected locations for the doublets.  The STIS feature are
Galactic {\FeII} absorption.}
\label{fig:limits} 
\end{figure*} 

G2 is associated with an extensive metal--line/LLS absorber at
$z=0.6601$ \citep{cat1,cat2,archiveI,ding05}, with $W_r({\Lya}) =
1.48$, $W_r({\Lyb}) = 0.84$, $W_r(2796) = 0.34$, $W_r(1548) = 0.32$,
and $W_r(1031) \leq 0.13$~{\AA} ($3~\sigma$).  The rotation curve of
G2 was presented by \citet{steidel02}, who showed that the bulk of the
{\MgII} absorbing gas kinematics are offset by $180$ {\kms} from the
G2 systemic velocity in the direction of galaxy rotation.  The {\CIV}
absorption spans the same velocity range as {\MgII}, though the
strongest component likely arises in a separate phase at the G2
systemic velocity \citep{cs03,ding05}.  As shown in
Figure~\ref{fig:limits}$a$, there is a remarkable complex of five
{\Lya} absorbers spanning the redshift of G1 in the G190H FOS
spectrum, first reported by \citet{cat2}.  Gaussian deblending
confirms five statistically significant {\Lya} components with
redshifts (rest--frame equivalent widths [{\AA}]) $z=0.66914$
($0.65\pm0.05$), $0.67157$ ($0.39\pm0.05$), $0.67355$ ($0.78\pm0.16$),
$0.67559$ ($0.79\pm0.20$), and $0.67707$ ($0.26\pm0.08$).  The total
equivalent width of the complex is $W_r({\Lya})=2.87$~{\AA}.  The
offset tick is the redshift of G1\footnote{We note that
\citet{chen01b} assign the $z=0.6716$ cloud to G1; however, the
equivalent widths they report appear to be misquoted within this
{\Lya} complex.}.

In Figure~\ref{fig:limits}$b$, we show the G160L FOS spectrum of the
Lyman limit for the {\Lya} complex encompassing G1 and G2.
\citet{cat1} measured $\tau_{_{LL}} \geq 3.5$, whereas
\citet{archiveI} obtained $\tau_{_{LL}} \geq 5.4$.  From this we infer
$\log N({\HI}) \geq 17.9$~[{\cmsq}].  However, simultaneous fitting to
the {\Lya}, {\Lyb}, and LL absorption at $z=0.6601$ yields $\log
N({\HI}) \simeq 18.3\pm0.3$.  We show this fit as the solid
curve in Figure~\ref{fig:limits}$b$.  We then modeled the Lyman limit
for the G1 {\Lya} complex to constrain the total $N({\HI})$.  We
tested cases that both included and excluded the LL break absorption
from the $z=0.6601$ system with various continuum extrapolations below
the LL break.  Applying the curve of growth to the {\Lya} equivalent
widths, we assigned equal Doppler $b$ parameters to all five
components and varied these $b$ parameters [$N({\HI})$ is dominated by
the $z=0.6736$ and $0.6756$ clouds].  We show four selected models
($b=30$, $35$, $40$, and $45$~{\kms}) as the short--dash--dot [total
G1 complex $\log N({\HI}) = 17.8$], long--dash [$16.9$], short--dash
[$16.3$], and dot--dot [$15.8$] curves, respectively.  Using the
$\chi^2$ statistic, we find that G1 {\Lya} complex has $\tau_{_{LL}}
\leq 0.39$ ($3~\sigma$) and estimate that no {\it single\/} component
can have $\log N({\HI})$ exceeding $\sim 16.5$.

In Figure~\ref{fig:limits}$c$, the expected regions for {\MgII},
{\CIV, {\NV}, and {\OVI} metal--line absorption are shown.  Ticks
provide the predicted locations of the doublet members at $z_{_{\rm
G1}}=0.6719$.  The regions cover $\pm 1200$~{\kms} (rest--frame)
relative to G1 over which {\it both\/} doublet members could be
detected ($0.6652 \leq z \leq 0.6785$, spanning the G1 {\Lya}
complex).  We obtained $3~\sigma$ rest--frame equivalent width limits
of 7~m{\AA}, 0.03~{\AA}, 0.11~{\AA}, and 0.21~{\AA} for {\MgII},
{\CIV}, {\NV}, and {\OVI}, respectively.  For {\MgII} and {\CIV},
$\log N({\MgII}) \leq 11.1$ and $\log N({\CIV}) \leq 12.9$,
independent of broadening.  Assuming $T=3\times10^5$ K, where the
{\OVI} fraction peaks in collisionally ionized gas, we find $\log
N({\NV}) \leq 13.8$ and $\log N({\OVI}) \leq 15.0$.

\section{Discussion}
\label{sec:discuss}

\subsection{On the Covering Factor}

Applying the $R^{\ast}$ and $\beta$ parameters quoted in
\S~\ref{sec:intro} to Equation~\ref{eq:rlb}, we obtain ``halo''
boundaries for G1 of $R(L) = 73$~kpc for $W_{r}(2796) \geq 0.3$~{\AA}
and $\tau_{_{\rm LL}} \geq 1$.  For $W_{r}(2796) \geq 0.02$~{\AA}, we
obtain $R(L) = 123$~kpc, and for $W_{r}(1548) \geq 0.17$~{\AA}, $R(L)
= 350$~kpc.  In Figure~\ref{fig:image}$a$, we illustrate the $R(L)$
boundaries for the LLS and {\MgII} thresholds. The solid circle
represents $W_r(2796)\geq 0.3$~{\AA} and $\tau_{_{\rm LL}} \geq 1$,
and the dash--dash circle represents $W_r(2796)\geq 0.02$~{\AA}.  The
$R(L)$ boundary for {\CIV} extends beyond the image section.  For G2,
we obtain $R(L) = 62$~kpc for $W_{r}(2796) \geq 0.3$~{\AA} and
$\tau_{_{\rm LL}} \geq 1$, and $R(L) = 228$~kpc for $W_{r}(1548) \geq
0.17$~{\AA}.

For G1 {\MgII} absorption, the $3~\sigma$ equivalent width upper limit
is a factor of $\simeq 40$ below $0.3$~{\AA} (the minimum expected
value) at the location $D/R(L) = 0.79$.  For weak {\MgII} absorption,
the limit is $\simeq 3$ times below the $0.02$~{\AA} sensitivity
threshold at $D/R(L) = 0.47$.  The limit for {\CIV} is a factor of
$\simeq 6$ below the $0.17$~{\AA} sensitivity threshold at $D/R(L) =
0.16$.  {\Lya} absorption with $W_r({\Lya}) \geq 0.30$~{\AA} is
expected within $D \leq 530$~kpc \citep{chen98,chen01b} and the
five--component G1 {\Lya} complex has total $W_r({\Lya}) = 2.87$~{\AA}.
However, at $D=58$~kpc, the expectation is that LLS absorption would
be detected in the ``halo'' of G1 and no clear feature is present to a
level suggesting $\tau_{_{\rm LL}} \leq 0.4$.  For G2, the conditions are
quite different.  {\MgII} and LLS absorption arises at $D/R(L) =
1.66$, well beyond the ``halo'' boundary, whereas the {\CIV}
absorption arises at $D/R(L) = 0.45$, per expectations.

There is now mounting evidence \citep[this
work;][]{bechtold92,cwc-china,tripp-china} that $C_f f_{gal} {\cal K}
< 1$ for {\MgII} absorption.  The very nature of selecting
absorber--galaxy pairs may have strongly contributed to previous
inferences that $C_f f_{gal} {\cal K} = 1$ and the applicability of
Equation~\ref{eq:rlb}.  Results for the G1 and G2 galaxies suggest
that, in some cases, this is true for {\CIV} and LLS absorption as
well.  A full assessment of the possible selection bias is highly
needed \citep[e.g.,][]{cc96}.  The application of a non--standardized
model of galaxy halos should now be incorporated into our appreciation
of galaxy halos in the context of $\Lambda$CDM formation models.

\subsection{Comparing the G1 and G2 Gas Properties}

In addition to the lack of metal--lines, the $\Delta v \simeq
1420$~{\kms} velocity spread of the G1 {\Lya} complex at $D=58$ kpc is
difficult to understand as a ``conventional'' gas halo.  On the
other hand, the G2 associated absorption at $D=103$ kpc has $\Delta v
\simeq 200$~{\kms} and is consistent with a photoionized halo,
including a optically thick phase ($T \simeq 8000$~K) and a diffuse
optically thin phase ($T \simeq 20,000 $~K) \citep[see Figure 6 and
Table 9 of][]{ding05}.  Accounting for $\log N({\HI}) = 18.3$
adopted in this work, the cloud metallicities from \citet{ding05}
could be as high as $\log (Z/Z_{\odot}) = -0.4$.

Adopting the solar abundance pattern of \citet{holweger01}, we
examined Cloudy \citep{ferland-cloudy} photoionization and collisional
\citep{sutherland93} models of the G1 {\Lya} complex in order to
constrain the metallicities.  For photoionization, the lack of [{\OII}]
emission (star formation) in G1 motivates an ultraviolet background
only model at $z=0.7$ \citep{haardt-madau96}.  Using the $N({\MgII})$
and $N({\CIV})$ limits, we computed a grid of $Z_{max} \equiv \log
Z/Z_{\odot}$, the upper limit on metallicity, for each {\Lya} cloud as
a function of ionization parameter in steps of $\Delta \log U = 0.2$
over the range $-3.5 \leq \log U \leq -0.5$ and $N({\HI}) \leq 16.5$,
determined from the {\Lya} curve of growth in steps of $\Delta b =
5$~{\kms}.


For {\MgII}, $Z_{max} \simeq \hbox{constant}$ for $\log U \leq -2.5$
for a given $N(\HI)$, and we find $Z_{max} \simeq \log
[14.4/N({\HI})]$.  At $-1.0 \leq \log U \leq -0.5$, where {\CIV}
peaks, $Z_{max} \simeq \log [12.5/N({\HI})]$.  Assuming $b_{_{\rm H}}
\simeq 40$~{\kms} \citep{williger06}, for the three lowest
$W_r({\Lya})$ clouds ($z=0.6771$, $0.6716$, and $0.6691$), the most
stringent limits are $Z_{max} \simeq -1.2$, $-1.6$, and $-2.6$ from
{\CIV}; $Z_{max}$ is less constrained with decreasing $U$ for $ \log U
\leq -1.0$ such that the $z=0.6771$ and $0.6716$ clouds can be
consistent with solar metallicity.  However, for the $z=0.6691$ cloud,
{\MgII} yields a hard limit of $Z_{max} \simeq -0.8$.  Assuming
$b_{_{\rm H}}=20$~{\kms} the three clouds have hard upper limits of
$Z_{max} \simeq -0.5$, $-1.5$, and $-1.8$ from {\MgII} for $\log U
\leq -2.5$.  An $\alpha$--group enhancement would decrease these hard
limits on $Z_{max}$ by as much as $0.3$--$0.5$ dex.  The most severe
constraints are $Z_{max} \simeq -1.6$, $-3.0$, and $-3.2$,
respectively, from {\CIV} for $-1.0 \leq \log U \leq 0.5$.

For the collisional ionization models, we
computed\footnote{$Z_{max}({\rm X}) = \log [ N_{lim}({\rm X})/N({\HI})
] + \log [ f({\HI})/f({\rm X})] - \log ( {\rm X}/{\rm H} ) _{\odot}$.}
$Z_{max}({\rm X})$ for each {\Lya} cloud as a function temperature (in
steps of $\Delta \log T = 0.05$~K) using the ionization fractions,
$f({\rm X})$, for {\HI}, {\CIV}, and {\OVI}. The $N({\CIV})$ and
$N({\OVI})$ limits and $N({\HI})$ are constrained by the curve of
growth as a function of temperature, assuming $b_{_{\rm X}}$ is
thermal.  The {\OVI} limit dominates for $T \geq 2\times 10^{5}$~K
[$b_{_{\rm H}} \geq 57$~{\kms}], and the {\CIV} limit dominates for
lower $T$ [$b_{_{\rm H}} \leq 57$~{\kms}].  At this transition $T$, the
hard limit is $Z_{max} = -0.6$ for the $z=0.6691$ cloud, and the hard
limit is $Z_{max} = -1.0$ for the two largest $W_r({\Lya})$ clouds
($z=0.6736$ and $0.6756$).  For the latter two clouds, if $T = 3
\times 10^{5}$ K, where {\OVI} peaks, then $\log N({\HI}) \simeq
14.7$ and $Z_{max} \simeq -2.8$; at $T = 10^{5}$~K, where
{\CIV} peaks, $\log N({\HI}) \simeq 16.0$ and $Z_{max} \simeq
-3.0$.

We consider the $\Lambda$CDM models of \citet{dave99} at $z\sim 1$ as
a guide to expectations for the gas properties in the proximity of G1
and G2.  At $D\simeq 60$~kpc (G1), the matter overdensities range from
$10 \leq \Delta \rho/\rho \leq 100$, where $\sim 80$\% of the
absorbers arise in shocked gas (accreting onto filaments or galaxy
halos) and $\sim 20$\% of absorbers arise in a photoionized,
adiabatically cooling, diffuse phase.  For $D\simeq100$ kpc (G2),
overdensities range from $\hbox{a few} \leq \Delta \rho/\rho \leq 30$,
where $\sim 60$\% arise in the shocked phase and $\sim 40$\% arise in
the diffuse phase.  

If the dominant clouds in the G1 {\Lya} complex are collisionally
ionized, the inferred properties are $T \sim 10^{5.0-5.5}$~K, $-1.0
\leq Z_{max} \leq -3.0$, $14.5 \leq \log N({\HI}) \leq 15.5$, and
$\Delta \rho/\rho \simeq 100$.  In $\Lambda$CDM cosmology, at $z\simeq
3$ this gas would have been in the photoionized diffuse phase with
$\Delta \rho/\rho \sim 10$ \citep{dave99,schaye01}.  The mean
metallicity of photoionized, diffuse {\Lya} forest clouds at $z\simeq
3$ is $[\hbox{C/H}]\simeq -2.5$ \citep{cowie98}, an enrichment level
consistent with the inferred $Z_{max}$ for collisional ionization in
the dominant clouds in the G1 {\Lya} complex at $z=0.7$.  A possible
scenario is that the dominant G1 clouds could have been enriched to
$[\hbox{C/H}]\simeq -2.5$ by $z=3$ while in the diffuse phase and
transitioned to shock heated gas by $z=0.7$, presumably due to their
accreting onto a filamentary structure in the vicinity of G1, without
having been further enriched by G1.  It is less likely that the
dominant clouds remained in the diffuse phase from $z=3$ to $z=0.7$
for which the \citet{dave99} and \citet{schaye01} models predict
$\Delta \rho/\rho \leq 10$, $\log N({\HI}) \simeq 14$, and $T \simeq
10^{4-4.3}$ K.  The dominant clouds cannot have this low $N({\HI})$
even for $b_{_{\rm H}} >100$~{\kms}, and even the weakest clouds would
have to be highly turbulent ($b_{_{\rm H}}>80$~{\kms}).
\citet{songaila96} reported that clustered {\Lya} lines at $z=3$ are
structured such that higher ionization clouds reside at the velocity
extremes and suggest that this is due to a layered structure to the
gas and is related to the physics of collapse.  A high resolution
spectrum of the Lyman series of the G1 {\Lya} complex could be very
informative regarding the physics underlying the gas.

The conditions of the G2 absorption is consistent with a diffuse phase
\citep{ding05}. A $\log N({\HI}) = 18.3$ corresponds to
$\Delta \rho/\rho > 1000$ at $z \leq 1$, that would have arisen from
an overdensity of $100$ at $z=3$ \citep{dave99,schaye01}.  This
implies the gas was already in the shocked phase by $z=3$.  It is
possible that the gas associated with the formation of G2 evolved more
rapidly than that of G1 and virialized by $z \simeq 3$.  If so, there
would be time for additional metallicity enrichment in this highly
evolved structure and for reprocessing from G2 to re--establish a
diffuse phase following the post--shock condensed phase.  The strong
{\OII} emission in G2 suggests star formation may have played a role,
perhaps inducing localized fountains and outflows.  This scenario is
analogous to the wind--driven gas observed in $z=3$ galaxies reported
by \citet{kurt05} and \citet{simcoe06}.

\section{Concluding Remarks}
\label{sec:conclude}

Even if the G1 and G2 gas structures extend $\simeq 3000$~{\kms}
between the galaxies \citep[see,][]{cat2}, inhomogeneities in the
overdensities on Mpc scales do not rule out the possible very
different evolution of the gas associated with the two galaxies, as
outlined above.  The presence of weak {\Lya} lines in the range $0.660
\leq z \leq 0.672$ would be highly suggestive of an {\HI} bridge
connecting the galaxies G1 and G2.  Both these two very dissimilar
absorber--galaxy pairs exhibit properties inconsistent with the
conventional model of galaxy halos (i.e., $C_f f_{gal} {\cal K} = 1$
and Equation~\ref{eq:rlb}) deduced from quasar absorption line
surveys.  Selection methods may have played an important role
\citep[see][]{cc96}; the notion that galaxy ``halos'' exhibit a level
of homogeneity in which they can be characterized by a standardized
covering factor, geometry, and boundary should be viewed critically.
We have suggested that the heterogeneous nature of extended gas
surrounding galaxies at impact parameters of several 10s to a 100
kiloparsecs may be strongly related to the range of overdensities from
which the galaxy and the gaseous environment evolved in the context of
$\Lambda$CDM models.

\acknowledgments

Support for this research through grant HST-AR-10644.01-A was provided
by NASA via the Space Telescope Science Institute, which is operated
by the Association of Universities for Research in Astronomy, Inc.,
under NASA contract NAS 5-26555.  G.G.K acknowledges support from
Sigma--Xi Grants in Aid of Research.  J.L.E. acknowledges support from
Research Cluster Fellowship from NMSU and a Graduate Student
Fellowship from NASA's New Mexico Space Grant Consortium.




{\it Facilities:} \facility{HST (WFPC--2, STIS, FOS)}, \facility{Keck I (HIRES, LRIS)}.

\newpage


\end{document}